\documentstyle[12pt]{article}
\begin{document}
\vskip 0.1in
\centerline{\Large\bf Holography in Closed Universes} 
\vskip .7in
\centerline{Dan N. Vollick}
\centerline{Department of Physics and Astronomy}
\centerline{Okanagan University College}
\centerline{Box 189}
\centerline{Salmon Arm, B.C.}
\centerline{V1E 4N3}
\vskip .9in
\centerline{\bf\large Abstract}
The holographic bound states that the entropy in a region cannot exceed one
quarter of the area (in Planck units) of the bounding surface. A version of the
holographic principle that can be applied to cosmological spacetimes has
recently been given by Fischler and Susskind. This version can be shown to
fail in closed spacetimes and they concluded that the holographic bound may
rule out such universes. In this paper I give a modified definition of the 
holographic bound that holds in a large class of closed universes. Fischler 
and Susskind also showed that the dominant energy condition follows from the
holographic principle applied to cosmological spacetimes with $a(t)=t^p$.
Here I show that the dominant energy condition can be violated by cosmologies
satisfying the holographic principle with more general scale factors.
\vskip 0.5in
\newpage
\section*{Introduction}
The holographic principle is a novel idea that was proposed by t'Hooft
\cite{tH1} and Susskind \cite{Su1} about 10 years ago. It states that a system 
in a given volume of space can be described by a theory on the boundary of that 
volume. In other words we can think of the degrees of freedom of the system as
residing on the boundary.
An explicit example of holography was first proposed by Maldacena 
\cite{Ma1}. He conjectured that string theory on AdS$_5\times S^5$ is
equivalent to $N=4$ supersymmetric Yang Mills theory on the boundary
of AdS$_5$.
One consequence of this principle is that the entropy in a given volume
cannot exceed one quarter of the area of the boundary of the volume. 
Throughout most of this paper (and much of the literature) the factor of
one quarter will be neglected.
   
The holographic principle was first applied to cosmology by Fischler and
Susskind \cite{Fi1}. They considered a region $V$ and proposed that the 
entropy of the matter that passed through the boundary of its past Cauchy
development, $\partial D^{-}(V)$, cannot exceed the area of its boundary 
$\partial V$. They showed that the holographic bound would be satisfied for a 
$k=0$ universe with $a(t)=t^p$ if the dominant energy condition holds. They 
also showed that their version of holography will fail in closed universes.
  
In this paper I give a modified version of the holographic principle that holds 
in a large class of closed universes. To motivate this modification let me
review the usual argument that leads to the holographic bound. Consider a
volume $V$ that is bounded by a surface of area $A$ and let $S>A/4$. 
Now add some 
energy to this region so that it forms a black hole of area $A$. 
Since the entropy of the
resulting black hole is $A/4$, this process would violate the second law of
thermodynamics. We therefore conclude that $S\leq A/4$.
This argument will fail in a closed universe if it is applied
to a region that is too large to collapse into a black hole before the
universe collapses into the final singularity. 
The condition that a region can collapse to form a
black hole before the final crunch can be approximated by the requirement that
$D^{+}(V)$ does not intersect the future singularity. In this
approximation $D^{+}(V)$ refers
to the original isotropic and homogeneous spacetime. If $D^{+}(V)$ does not
intersect the future singularity then every particle in $V$ will cross
$\partial D^{+}(V)$. The modified holographic bound proposed in this paper
can be stated as follows: the entropy of the matter in a volume $V$ that
passes through both $\partial D^{-}(V)$ and $\partial D^{+}(V)$ cannot exceed
the area of $\partial V$. Note that this proposal reduces the the Fischler and
Susskind proposal in open universes since all the matter in $V$ will pass 
through $D^{+}(V)$.
   
I also examine the relationship between the holographic principle and the
dominant energy condition. Fischler and Susskind showed that the holographic
bound implies the dominant energy condition for spacetimes with $a(t)=t^p$.
Here I show that this is not the case
for more general scale factors.

\section*{Spatially Flat Universes}
In this section the results of Fischler and Susskind \cite{Fi1} will be 
reviewed.
The spacetime will be taken to be a d+1 dimensional FRW universe with metric
\begin{equation}
ds^2=-dt^2+a(t)^2dx^kdx_k\; ,
\end{equation}
where $k=1...d$. 
  
Consider a spherical region $V$ at time $t$. If this region is smaller than 
the particle horizon its past Cauchy development $D^-(V)$ will form a
cone with its tip in the chronological future of the past singularity. If the 
region is the same size as the particle horizon $D^-(V)$ is a cone with its
tip at the past singularity and if the region is larger than the horizon
$D^-(V)$ is a truncated cone. Fischler and Susskind proposed that
the entropy of the matter that passed through $\partial D^-(V)$ cannot
exceed the area of $\partial V$. Thus, the total entropy in a region smaller 
than or equal to the horizon size cannot exceed the area of that region.
  
Since the entropy per unit coordinate volume $\sigma$ is constant the entropy 
in a spherical region of coordinate radius $R_H$ is given by
\begin{equation}
S=a_d\sigma R_H^d \; , 
\end{equation}
where $a_d=\pi^{d/2}/\Gamma(d/2+1)$ and
$R_H$ is the horizon size in comoving coordinates which is given by
\begin{equation}
R_H(t)=\int_0^t\frac{dt^{'}}{a(t^{'})} .
\end{equation}
Here I will assume that this integral is finite.
The physical area of the bounding sphere is
\begin{equation}
A=b_d\left[ aR_H\right]^{d-1} ,
\end{equation}
where $b_d=2\pi^{d/2}/\Gamma(d/2)$.
Thus, the holographic bound is satisfied iff
\begin{equation}
a_d\sigma R_H^d\leq b_d\left[ aR_H\right]^{d-1} .
\end{equation}
From this we deduce that
\begin{equation}
\sigma a_d\int_0^t\frac{dt'}{a(t')}\leq b_da(t)^{d-1}.
\label{hol1}
\end{equation}
This must hold at all times greater than the Planck time.
  
In a $k=0$ FRW universe the scale factor satisfies
\begin{equation}
\left( \frac{\dot{a}}{a}\right)^2=\frac{8\pi G}{3}\rho,
\end{equation}
where $\rho$ is the matter density. This implies that if the universe is 
initially expanding it will continue to expand ($\rho>0$)
and we see that $a(t)$ is a
monotonically increasing function of $t$. Thus, $1/a(t)$ is a 
monotonically decreasing function of $t$ and we have
\begin{equation}
\int_0^t\frac{dt'}{a(t')} > \frac{t}{a(t)} .
\end{equation}
Using (\ref{hol1}) we find that the holographic bound implies that
\begin{equation}
a(t) > \left[\left(\frac{\sigma a_d}{b_d}\right) t\right]^{1/d} .
\label{hol2}
\end{equation} 
It is important to note that the holographic bound implies this inequality
but not vice versa. Thus, if the above bound is violated the holographic 
bound will also be violated.
  
If $a(t)=t^p$ equation (\ref{hol1}) gives $p\geq 1/d$, as obtained by Fischler
and Susskind. It is convenient to define $\eta$ by
\begin{equation}
\eta =\frac{S}{A}\sim t^{1-dp}\; .
\end{equation}
We therefore require that $p\geq 1/d$ for $\eta$ to be small at late times. 
Note however
that if $p>1/d$ then $\eta$ will diverge as $t\rightarrow 0$. This does not
present a problem if $\eta\stackrel{<}{\sim} 1$ at the Planck time $t_p$. 
Present estimates
give
\begin{equation}
\eta (t_d)\sim 10^{-28}
\end{equation}
in Planck units at decoupling. 
Assuming that the universe is radiation dominated for $t<t_d$ and taking
$d=3$ gives
\begin{equation}
\eta(t)\sim 10^{-28}\left[\frac{t_d}{t}\right]^{1/2} .
\end{equation}
Since $(t_d/t_p)^{1/2}\sim 10^{28}$ we see that 
$\eta\sim 1$ for $t>t_p$.
  
Fischler and Susskind also showed that for a power law expansion
the holographic principle constrains the velocity of sound to be less than or 
equal to c. Their argument is as follows. For $P=\gamma\rho$ the Einstein field 
equations give
\begin{equation}
a(t)\sim t^p
\end{equation}
with $p=2/d(1+\gamma)$. The holographic bound then gives $|\gamma| \leq 1$,
which is the usual constraint that the speed of sound does not exceed c.
This can also be stated in terms of the dominant energy condition, which
constrains the speed of the flow of energy to be less than or equal to the 
speed of light \cite{Wa}, as follows. For $a(t)=t^p$ the holographic bound 
implies the dominant energy condition.
  
It is possible however to violate the dominant energy condition without 
violating the holographic bound. Consider the constraint (\ref{hol1})
imposed by holography and let $a_0(t)$ satisfy the strict inequality. 
Now consider a scale factor $a(t)$ given by
\begin{equation}
a(t)=a_0(t)+\epsilon(t)
\end{equation}
with $|\epsilon(t)|<<a_0(t)$ for $t>0$ and 
$|\dot{\epsilon}(t)|>>|\dot{a}_0(t)|$, 
$|\ddot{\epsilon}(t)|>>|\ddot{a}_0(t)|$ on some interval $I$. For 
$\epsilon(t)$ sufficiently small $a(t)$ will satisfy (\ref{hol1}). The
pressure and density on $I$ are given by
\begin{equation}
\rho\simeq\frac{3\dot{\epsilon}^2}{\kappa a_0^2}
\end{equation}
and 
\begin{equation}
P\simeq -\frac{1}{\kappa}\left[\frac{2\ddot{\epsilon}}{a_0}+
\frac{\dot{\epsilon}^2}{a_0^2}\right]
\end{equation}
where $\kappa=8\pi G$. To create violations of the dominant energy condition 
consider solutions with $P>\rho$. This implies that
\begin{equation}
a_0\ddot{\epsilon}+2\dot{\epsilon}^2<0 .
\label{ineq}
\end{equation}
Thus, any function $\epsilon(t)$ that satisfies this inequality on some
time interval $I$ will violate the dominant energy condition on that interval.
  
For example let $a_0(t)=t^p$ and $\epsilon(t)=\alpha\sin(\beta t)$ and take
\begin{equation}
\lim_{t\rightarrow 0}\frac{\epsilon(t)}{a_0(t)}=0\; .
\end{equation}
This implies $p<1$. We also take $p>0$ so that $a\rightarrow 0$ as $t
\rightarrow 0$.
The maximum value of $\epsilon/a_0$ is
\begin{equation}
\left(\frac{\epsilon}{a_0}\right)_{max}=\alpha\beta^p\left[\frac{\sin[\Lambda(p)]}
{\Lambda(p)^p}\right]
\end{equation}
where $\Lambda(p)$ satisfies $\Lambda-p\tan\Lambda=0$ and is a 
monotonically decreasing function of $p$.
In fact $\Lambda(p)$ 
can be approximated by $\Lambda (p)\simeq 
1.6\sqrt{1-p}$ for $0<p<1$. Since
\begin{equation}
0<\frac{\sin[\Lambda(p)]}{\Lambda(p)^p}<1
\end{equation}
for $0<p<1$, we require that $\alpha\beta^p<<1$. Now
\begin{equation}
\frac{\dot{\epsilon}}{\dot{a}_0}=\frac{\alpha\beta}{p}t^{1-p}\cos(\beta t)
\end{equation}
and
\begin{equation}
\frac{\ddot{\epsilon}}{\ddot{a}_0}=\frac{\alpha\beta^2}{p(1-p)}t^{2-p}
\sin(\beta t).
\end{equation}
Thus, at sufficiently late times $|\dot{\epsilon}|>>\dot{a}_0$ and
$|\ddot{\epsilon}|>>|\ddot{a}_0|$, if $\sin(\beta t)$ or $\cos(\beta t)$ are 
not too close to zero.
The inequality (\ref{ineq}) becomes
\begin{equation}
t^p\sin(\beta t)-2\alpha\cos^2(\beta t)>0 .
\end{equation}
for $\alpha>0$.
This will certainly be satisfied 
at sufficiently late times when $\sin(\beta t)
>0$. Thus, the holographic bound
does not imply the dominant energy condition. 
\section*{Closed Universes}
The metric of a closed 3+1 dimensional closed universe can be written as
\begin{equation}ds^2=-dt^2+a(t)^2\left( d\chi^2+\sin^2\chi d\Omega^2\right) .
\end{equation}
The scale factor will be taken to satisfy $a(0)=0$ and $a(T)=0$ with $T>0$.
Closed universes that do not collapse into a final singularity 
will be discussed at the end of this
section. The scale factor satisfies
\begin{equation}
\left(\frac{\dot{a}}{a}\right)^2=\frac{8\pi G}{3}\rho(a)-\frac{1}{a^2}
\end{equation}
and it is easy to see that $a(t)$ will have its maximum value at $t=T/2$
and will be symmetric about this value.
The comoving particle horizon is given by
\begin{equation}
\chi_H(t)=\int_0^t\frac{dt^{'}}{a(t^{'})}
\label{chi1}
\end{equation}
and $\eta=S/A$ is given by
\begin{equation}
\eta=\sigma\left[\frac{2\chi_H-\sin(2\chi_H)}{4a^2\sin^2\chi_H}\right] .
\label{eta}
\end{equation}
In the early universe where $\chi_H<<1$ we have
\begin{equation}
\eta\simeq\frac{\sigma\chi_H}{3a^2(\chi_H)}
\label{c1}
\end{equation}
This will diverge as $t\rightarrow 0$
if $a(t)\sim t^p$ for $p>1/3$. At the Planck time we
require that
\begin{equation}
\frac{\chi_H(t_p)\sigma}{3a^2(t_p)}\stackrel{<}{\sim} 1 .
\label{c2}
\end{equation}
The causally connected region becomes the entire sphere as $\chi_H\rightarrow
\pi$. If this is reached at time $T$ (the big crunch)
then $\eta$ at a Planck time before the
big crunch is given by
\begin{equation}
\eta(T-t_p)\simeq\left[\frac{\pi\sigma}{2a^2(t)(\pi-\chi_H(t))^2}
\right]_{t=T-t_p}.
\label{eta1}
\end{equation}
Thus, from (\ref{c1}) and (\ref{eta1})we see that 
 $\eta(T-t_p)>>1$ unless $\eta(t_p)<<1$,
which is not the case for our universe.
 
If $\chi_H=\pi$ at some value of $t<T$ then 
$\eta$ will diverge before the big crunch 
and the holographic principle will be violated. Finally, if $\chi_H$ does not
reach $\pi$ before the universe recollapses then $\eta\sim \sigma/a^2$ and $\eta
>>1$ at a Planck time before the end unless $\eta(t_p)<<1$. Thus, the 
holographic principle seems to run into trouble in a closed universe.
  
To avoid the above problem it is necessary to reformulate the holographic
principle. Fischler and Susskind assumed that only the matter
in a volume V that passed through $\partial D^{-}(V)$ can be included in the
holographic bound. Thus, the holographic bound can only be applied to all
the entropy in a region if the region is horizon sized or smaller.
  
I propose the following modification of the above idea. Only the entropy of the
matter that passes through both $\partial D^{-}(V)$ and $\partial D^{+}(V)$
can be included in the holographic bound. This is equivalent to the 
Fischler and Susskind proposal in open universes, since all the particles in 
a volume $V$ will pass through $\partial D^{+}(V)$. However, in a closed 
universe $D^{+}(V)$ may be truncated on the future
singularity and not all of the particles in $V$ will pass through $\partial
D^{+}(V)$. This modification implies that we cannot apply the holographic
principle to the total entropy in a region if its future Cauchy development
intersects the future singularity.
  
One possible justification for this proposal follows from the argument
that is used to support the holographic bound. Consider a volume of space $V$ 
that is bounded by a surface area $A$ and let $S>A/4$. Now add some energy to 
this region so that it forms a black hole of surface area $A$. 
Since the entropy of the black hole 
is $A/4$ this process will violate the second law of thermodynamics (see Wald
\cite{Wa1}, for an objection to this argument). Thus,
$S\leq A/4$. However, this argument will
fail in a closed universe for large regions at late times, as there will
be insufficient time for a black hole to form. Thus, we can apply the
holographic bound $S/A<1$ only to regions $V$ that could collapse to form
a black hole before the big crunch. For an observer outside the collapsing
matter the time scale for the final approach to the black hole state is of
order of the light crossing time, $\sim R_s/c$, where $R_s$ is the Schwarzschild
radius. Thus, an approximate requirement for the formation of the black hole
is that $D^{+}(V)$ does not intersect the future singularity. In this 
approximation $D^{+}(V)$ refers to the future Cauchy development of the original
homogeneous and isotropic universe, i.e. without the black hole.
The intersection of the causal past of a point on the future
singularity with the constant time hypersurface will be some volume, say
$\tilde{V}_H$. If $V_H$ denotes a horizon sized volume it will be the minimum
value of $\tilde{V}_H$ and $V_H$ that gives
the largest region that the holographic principle can be applied to on that
hypersurface. 
   
By symmetry if $t<T/2$ the largest volume that can be used will be $V_H$
and if $t>T/2$ the largest volume that can be used will be $\tilde{V}_H$.
In fact we only need to consider $\eta$ on $0<t\leq T/2$. If $\eta<1$ on
this interval the holographic bound will be satisfied at all times.
It is useful to define $\chi_{max}$ by $\chi_{max}=\chi(T)$. 
If $\chi_{max}\geq 2\pi$ we have $\chi_H(T/2)\geq \pi$.
In this case $\eta$ diverges at $t\leq
T/2$ and the holographic bound will be violated.
  
If $\chi_{max}<2\pi$ then $\chi_H(T/2)<\pi$ and we
avoid the above divergence. This does not guarantee however that $\eta<1$
for $0<t<T/2$. To see if $\eta<1$ one needs to know $a(\chi_H)$.
If $\chi_{max}$ is not too close to $2\pi$ then
\begin{equation}
\eta\sim\frac{\sigma\chi_H}{3a^2}
\end{equation}
on $0<t<T/2$
and the holographic bound will be satisfied for $\eta(t_p)\stackrel{<}{\sim}1$
if $a$ increases at least as fast as $\sqrt{\chi_H}$.
  
I will examine a radiation dominated universe and show that if $\eta<1$ in the
early universe the holographic bound will be satisfied at all times. In a 
matter dominated universe $\chi_{max}=2\pi$ so that $\eta$ will diverge as
$\chi_H\rightarrow \pi$. As discussed earlier, one could argue that using 
$D^{+}(V)$ is only an approximate implementation of the holographic principle.
If one takes this viewpoint it is possible that the holographic bound is not
violated in matter dominated universes since $\chi_H$ just makes it to $2\pi$.
A more detailed analysis would be required to resolve the issue. 
   
Equation (\ref{chi1}) for $\chi_H$ can be written as
\begin{equation}
\chi_H(a)=\int_0^a\frac{da}{a\dot{a}}=\int_0^a\frac{da}{\sqrt{\frac{8\pi G}
{3}\rho a^4-a^2}}\; ,
\label{chia1}
\end{equation}
for $0<t\leq T/2$. If $T/2<t<T$ equation (\ref{chi1}) becomes
\begin{equation}
\chi_H(a)=\frac{\chi_{max}}{2}+\int_a^{a_{max}}\frac{da}
{\sqrt{\frac{8\pi G}{3}\rho
a^4-a^2}} .
\label{chia2}
\end{equation}
In a radiation dominated universe $\rho=\rho_0/a^4$, where $\rho_0$ is a 
constant.
Equations (\ref{chia1}) and (\ref{chia2}) can be easily integrated to give
\begin{equation}
a(\chi_H)=\sqrt{\tilde{\rho}}\sin(\chi_H),
\end{equation}
where $\tilde{\rho}=8\pi G\rho_0/3$. From this expression it is easy to see
that $\chi_{max}=\pi$. From (\ref{eta}) we have
\begin{equation}
\eta=\sigma\left[\frac{2\chi_H-\sin(2\chi_H)}{4\tilde{\rho}\sin^4(\chi_H)}
\right]
\end{equation}
where we consider $0<\chi<\pi/2$. In the early universe $\eta$ is given by
\begin{equation}
\eta\simeq\frac{\sigma}{3\tilde{\rho}\chi_H},
\end{equation}
so that $\eta$ diverges as $\chi_H\rightarrow 0$. As $\chi_H$ increases $\eta$
decreases until $\chi_H\simeq 1$ and $\eta\simeq 0.5\sigma/\tilde{\rho}$. 
The value of
$\eta$ then slowly increases to $\eta=\frac{\pi\sigma}{4\tilde{\rho}}$ at
$\chi_H=\pi/2$. Thus, if $\eta\stackrel{<}{\sim}1$ at the Planck time the
holographic bound will be satisfied at all times.
   
Finally, let me make a few brief comments on closed universes that do not
collapse into a final singularity, but instead expand forever. A simple example 
of such a universe is a Lemaitre spacetime, which is a dust filled universe
with a positive cosmological constant. In this case the holographic bound
proposed in this paper reduces to the bound of Fischler and Susskind. If
\begin{equation}
\chi_{max}=\int_0^{\infty}\frac{dt}{a(t)}\geq\pi
\label{chimax}
\end{equation}
then the holographic bound will certainly be violated (see equation 
(\ref{eta})). On the other hand, if $\chi_{max}<\pi$ the holographic bound
may not be violated and one has to check for the given $a(t)$. An example of a 
universe in which $\chi_{max}$ is bounded even though the universe expands
forever is a Lemaitre universe. In the early universe $a(t)\sim t^{2/3}$ so 
that (\ref{chimax}) is bounded at the lower limit of integration and in the
late universe $a(t)\sim e^{Ht}$ so that (\ref{chimax}) is bounded at the
upper limit of integration.
   
It is interesting to note that the version of holography 
proposed in this paper requires 
knowledge of the entire spacetime. Whether or not one can apply the
usual statement that $S/A<1$ for a given region depends on the future
evolution of the spacetime.
\section*{Conclusion}
In this paper I proposed a modification of the Fischler and Susskind entropy
bound. In their proposal the entropy of the matter in a volume $V$ that passes
through $\partial D^{-}(V)$ cannot exceed the area $\partial V$. It is easy to
show that this bound fails in a closed universe. In my proposal the matter has
to pass through both $\partial D^{-}(V)$ and $\partial D^{+}(V)$ to be included
in the holographic bound. This proposal reduces to the bound of
Fischler and Susskind in open universes since all the particles in $V$ will
pass through $\partial D^{+}(V)$. I showed that the holographic bound
proposed in this paper holds in a large class of closed universes, including
radiation dominated universes.
  
I also examined the relationship between the holographic bound and the
dominant energy condition. Fischler and Susskind showed that the holographic 
implies the dominant energy condition in open universes with $a(t)\sim t^p$.
Here I showed that this is not the case for more general scale factors.

\end{document}